\documentstyle[psfig,12pt]{article}

\title{Level spectroscopy II -- application to the Gaussian model --}

\author{Kiyohide Nomura	and Atsuhiro Kitazawa	\\
        {\em Department of Physics,} \\
        {\em Tokyo Institute of Technology,} \\
        {\em Tokyo 152, Japan.} \\
}

\date{\today}

\begin{document}

\maketitle

\begin{abstract}
It was a difficult problem to determine the Gaussian fixed line from
the numerical data, because close to the Berezinskii-Kosterlitz-Thouless 
multicritical point 
the divergence of the correlation length becomes very slow.  
Considering the renormalization group behavior, we find an efficient
method to determine the Gaussian fixed line.  
\end{abstract}

PACS numbers: 75.40.Cx, 05.70.Fh, 11.10.Hi, 75.10.Jm

\pagebreak

The two dimensional (2D) sine-Gordon model, 
which is defined with the Lagrangian
\begin{equation}
        {\cal L}={1 \over 2\pi K} (\nabla \phi)^2 
        +{y_1 \over 2\pi \alpha^2 } \cos \sqrt{2} \phi,
        \label{eqn:SGlag}
\end{equation}
(with the identification $\phi \equiv \phi + 2\pi /\sqrt{2}$) 
and the dual field $\theta$
\begin{equation}
	\partial_x \phi = -\partial_y (iK \theta), 
	\partial_y \phi = \partial_x (iK \theta), 
\end{equation}
plays an important role in 2D classical and 1D quantum systems, 
such as the 2D XY model, 2D helium,
1D quantum spin models and 1D electron models.  
It is also related with 
the ${\rm U(1) \times Z_2 \times Z_2}$ conformal field theory 
or the Tomonaga-Luttinger liquid.   
The sine-Gordon model has the Berezinskii-Kosterlitz-Thouless (BKT) 
\cite{Berezinskii,Kosterlitz-T,Kosterlitz} 
phase boundaries and the Gaussian fixed line.  

However, it was a difficult problem to determine the phase boundary 
and the universality class of the sine-Gordon model 
from the numerical calculation.  Finite-size
scaling technique, which is an efficient method to analyze the
ordinary second order transition, leads to false conclusions for
the BKT transition \cite{Bonner-M,Solyom-Z,Seiler-S,Edwards-G}, 
since the divergence of the correlation length 
is very slow $\xi \sim \exp ( const./\sqrt{T-T_c})$ 
and there are logarithmic corrections caused by the marginally
irrelevant field.  
One of us \cite{Nomura} has proposed a remedy for these problems, 
the level spectroscopy, based on the renormalization analysis 
and the ${\rm SU(2)/Z_2}$ symmetry of the BKT transition.  

About the Gaussian fixed line, in the neighborhood of the BKT
multicritical point, since the divergence of the correlation length 
is very slow, it is difficult to determine the critical line with
the finite-size scaling technique.  
In addition, from the simple finite-size scaling analysis of 
the Gaussian model, two (pseudo) fixed points or no fixed
point are obtained, 
which causes a finger-like massless region around the true
Gaussian fixed line \cite{Botet-J-K,Glaus-S,Schulz-Z,Solyom-Z}.  
In this letter we propose an improved method to obtain the Gaussian
fixed line.  

The renormalization group equations for (\ref{eqn:SGlag}) 
are \cite{Kosterlitz}
\begin{eqnarray}
        \frac{d K^{-1} (l)}{dl}&=& \frac{1}{8}y_1^2 (l), \nonumber\\
        \frac{dy_1 (l)}{dl}&=& \left( 2-\frac{K(l)}{2} \right) y_1(l).
        \label{eqn:renormal}
\end{eqnarray}
For the finite system, $l$ is related to $L$ by $e^l = L$.  
Close to the BKT multicritical point $K\approx 4, |y_1| \ll 1$, 
the divergence of the correlation length close to the Gaussian line
becomes very slow, as mentioned above.

On the Gaussian fixed line ($y_1=0$), the operators 
$O_{n,m} \equiv \exp (i n \sqrt{2} \phi + i m \sqrt{2} \theta)$ 
has a critical dimension $x_{n,m}$ and a spin $l_{n,m}$ 
given by \cite{Kadanoff-Brown}
\begin{equation}
        x_{n,m}=\frac{1}{2} \left( n^2 K +\frac{m^2}{K} \right), \:
        l_{n,m}= nm.
\end{equation}
When we denote the transfer matrix of a strip of width $L$ with
periodic boundary condition by $\exp(-H)$, then the energy gaps
$\Delta E_{n,m}$ are related to the scaling dimension as 
\cite{Cardy84} 
\begin{equation}
	\Delta E_{n,m} (L) = \frac{2 \pi v x_{n,m}}{L}, 
\end{equation}
where $v$ is the ``velocity of light''.  
Therefore, we obtain
\begin{equation}
	\Delta E_{0,2} (L) /\Delta E_{0,1} (L) = 4.
\end{equation}

On the other hand, the off-critical behavior $y_1 \neq 0, K<4$
is considered as follows.  In this case eq. (\ref{eqn:renormal}) is
infrared unstable and the correlation length $\xi$ becomes finite.  
In the  $L \gg \xi$ limit, it is considered that the excitation of $m=2$
is a scattering state of two excitations of $m=1$, which means 
\begin{equation}
	\Delta E_{0,2} (L) /\Delta E_{0,1} (L) = 2.
\end{equation}
Therefore, we expect that the ratio 
$\Delta E_{0,2}(L)/\Delta E_{0,1}(L)$ becomes maximum at $y_1=0$ 
and decreases as increasing $|y_1|$ (see Figure 1).  

Consider the other type ratio
$
	\Delta E_{0,2} (2L) /\Delta E_{0,1} (L)
$.  
It takes 2 at the Gaussian line, and in the massive region 
it also takes 2, therefore it is expected
\begin{equation}
	\Delta E_{0,2} (2L) /\Delta E_{0,1} (L) =2
	\label{eqn:invariant}
\end{equation}
independently of $y_1$ (see Figure 2).  

We shall explain these behaviors with the renormalization group argument.  
The renormalized critical dimensions are \cite{Giamarchi-S}
\begin{equation}
	x_{0,m} = \frac{m^2}{2}K^{-1}(l)
\end{equation}
However, with only these relations, we cannot explain the behavior 
$\Delta E_{0,2}(L)/\Delta E_{0,1}(L)$ close to the Gaussian line, 
since the ratio of $x_{0,m}$ is independent of $y_1$.  
Considering the renormalization constant $L_0$, 
which is included in the definition $l$ as $l \equiv \log (L/L_0)$, 
we can understand these features (6),(7),(8).  
In order to satisfy the relation (\ref{eqn:invariant}), 
the renormalization constant for the $\Delta E_{0,2}$ should be 
$2 L_0$, where $L_0$ is the renormalization constant for 
$\Delta E_{0,1}$.  Then, we obtain 
\begin{eqnarray}
	\frac{\Delta E_{0,2} (L)}{\Delta E_{0,1} (L)}
	& = & 4 \left[ \frac{K^{-1}(l-\log 2)}{K^{-1}(l)}
	\right]_{l=\log L/L_0}	\nonumber\\
	& \approx &
	\frac{4}{K^{-1}(l)} \left[K^{-1}(l)-\log 2 \frac{d K^{-1}(l)}{dl} 
	\right]_{l=\log L/L_0}
	\nonumber\\
	& = &
	4 \left[1 - K(l)\frac{\log 2}{8}
	y_1^2(l) \right]_{l=\log L/L_0},
\end{eqnarray}
by using eqs.(\ref{eqn:renormal}).  
Consequently we can determine the Gaussian fixed line 
from the extremum point of the  $\Delta E_{0,2}(L)/\Delta E_{0,1}(L)$.  

As a physical model, we study the S=1 bond-alternating XXZ model 
\cite{Kitazawa-N-O} 
\begin{equation}
	H = \sum_{j=1}^L (1 + \delta (-1)^j)
	( S^x_j S^x_{j+1}+S^y_j S^y_{j+1}+ \Delta S^z_j S^z_{j+1}).
\end{equation}
In this case $\Delta E_{0,m}(L)$ corresponds to $\Delta E (S^z_T=m,L)$, 
where $S^z_T = \sum S^z_j$.  
In Figure 1, we show the ratio $\Delta E (2, L)/ \Delta E(1,L)$.
And in Figure 2, we show the ratio $\Delta E (2, 2L)/ \Delta E(1,L)$.
Their behaviors are consistent with the renormalization discussion.  
The remaining correction $1/L^2$ can be explained by 
the irrelevant operator $L_{-2} \bar{L}_{-2} {\bf1 }$ 
with the critical dimension $x=4$.  

Here we analyze the corrections caused 
by the irrelevant operator $L_{-2} \bar{L}_{-2} {\bf1 }$.   
For simplicity we treat the case $|y_1| \ll |K-4|$.  
In this case $y_1 = y_1(0) L^{2 - K/2}$.  
And we assume the correction from the $x=4$ term as 
$(c_1 +c_2(\delta-\delta_c))L^{-2}$ and the dependence of $y_1$ 
on the parameter $\delta$ as 
$y_1(0) = c_3(\delta-\delta_c)$.  
Then the extremum of the $\Delta E_{0,2}(L)/\Delta E_{0,1}(L)$ is
situated at 
\begin{equation}
	\delta-\delta_c 
	=  \frac{1}{K \log 2}\frac{c_2}{c_3^2} L ^{K-6},
	\label{eqn:correction}
\end{equation}
which rapidly converges to 0 in the $L \rightarrow \infty$ limit.  

Finally we criticize the method proposed in \cite{Nomura}.  
In this method, there appears the term
$
	y_\phi^2 (l)/ (K(l)-4)
$.  
However, close to the BKT multicritical point, the sign of $K(l)-4$ 
changes, therefore it becomes difficult to determine the maximum.  

\bigskip

The authors thank the Supercomputer Center, Institute for
Solid State Physics, University of Tokyo for the facilities
and the use of the FACOM VPP500.

\begin{figure}
\begin{center}
\psfig{figure=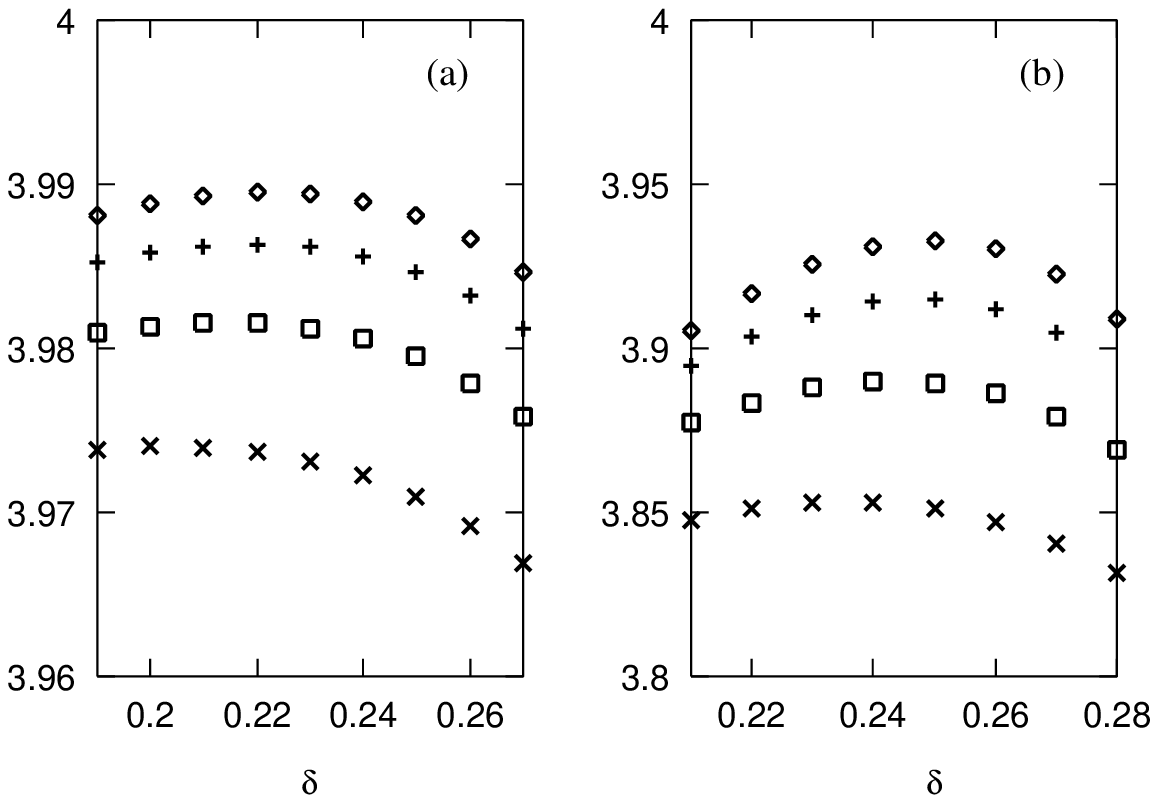,width=9cm}
\end{center}
\caption{
Ratio of excitation energies $\Delta E(2,L)/\Delta E(1,L)$ for 
$L=10$ ($\times$), $L=12$ ($\Box$), $L=14$ ($+$) and $L=16$ ($\Diamond$). 
(a) $\Delta=0$ where the critical value of $\delta$ is $\delta = 0.23$.
(b) $\Delta=0.5$ where the critical value of $\delta$ is $\delta = 0.25$.
}

\bigskip

\begin{center}
\psfig{figure=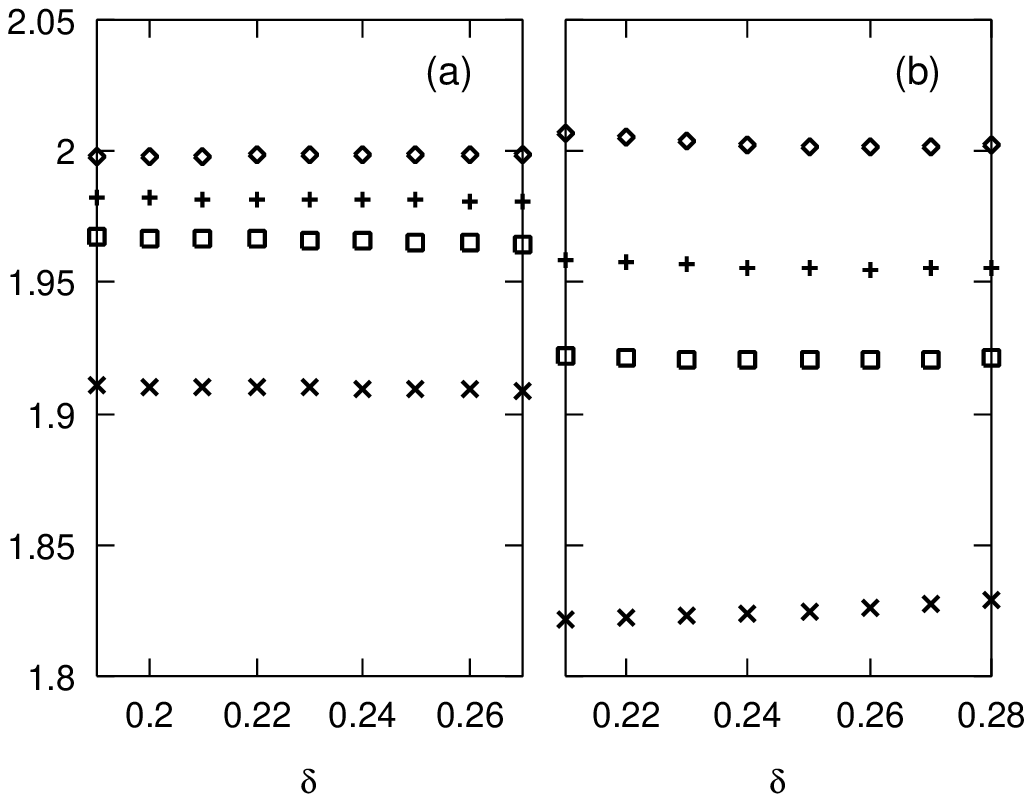,width=9cm}
\end{center}
\caption{
Ratio of excitation energies  $\Delta E(2,2L)/\Delta E(1,L)$ for 
$L=4$ ($\times$), $L=6$ ($\Box$) and $L=8$ ($+$). 
$\Diamond$ is the extraporated value.
(a) $\Delta=0$. (b) $\Delta=0.5$.
}
\end{figure}

\end{document}